\documentclass{emulateapj}
\usepackage{rotating}

\def\lsim{\,\lower2truept\hbox{${<\atop\hbox{\raise4truept\hbox{$\sim$}}}$}\,}
\def\gsim{\,\lower2truept\hbox{${> \atop\hbox{\raise4truept\hbox{$\sim$}}}$}\,}

\shorttitle{Polarization of the WMAP Point Sources}
\shortauthors{L\'opez-Caniego et al.}

\begin{document}

\title{Polarization of the WMAP Point Sources}

\author{M. L\'opez-Caniego}
\affil{Instituto de F\'isica de Cantabria (CSIC-UC), Santander, Spain 39005}
\affil{Astrophysics Group, Cavendish Laboratory, J.J. Thomson Avenue, Cambridge, United Kingdom  CB3 0E1}
\email{caniego@ifca.unican.es}

\author{M. Massardi} 
\affil{INAF-Ossevatorio Astronomico di Padova, Padova, Italy 35122}

\author{J. Gonz\'alez-Nuevo} 
\affil{SISSA-I.S.A.S, Trieste, Italy 34014}

\author{L. Lanz}
\affil{Instituto de F\'isica de Cantabria (CSIC-UC), Santander, Spain 39005}
\affil{Departamento de F\'isica Moderna, Universidad de Cantabria, Santander, Spain 39005}

\author{D. Herranz} 
\affil{Instituto de F\'isica de Cantabria (CSIC-UC), Santander, Spain 39005}

\author{G. De Zotti} 
\affil{INAF-Ossevatorio Astronomico di Padova, Padova, Italy 35122}
\affil{SISSA-I.S.A.S, Trieste, Italy 34014}

\author{J.L. Sanz} 
\affil{Instituto de F\'isica de Cantabria (CSIC-UC), Santander, Spain 39005}

\and

\author{F. Arg\"ueso} 
\affil{Departamento de Matem\'aticas, Universidad de Oviedo, Oviedo, Spain 33007}

\begin{abstract}
  The detection of polarized sources in the WMAP 5-year data is a very
  difficult task. The maps are dominated by instrumental noise and
  only a handful of sources show up as clear peaks in the $Q$ and $U$
  maps.  Optimal linear filters applied at the position of known
  bright sources detect with a high level of significance a polarized
  flux $P$ from many more sources, but estimates of $P$ are liable to
  biases. Using a new technique, named the \emph{filtered fusion
    technique}, we have detected in polarization, with a significance
  level greater than 99.99\% in at least one WMAP channel, 22 objects,
  5 of which, however, do not have a plausible low radio frequency
  counterpart and are therefore doubtful. Estimated polarized fluxes
  $P < 400\,$mJy at 23 GHz were found to be severely affected by the
  Eddington bias. The corresponding polarized flux limit for
  Planck/LFI at 30 GHz, obtained via realistic simulations, is 300
  mJy.  We have also obtained statistical estimates of, or upper
  limits to the mean polarization degrees of bright WMAP sources at
  23, 33, 41, and 61 GHz, finding that they are of a few percent.

\end{abstract}


\section{Introduction} \label{sec:intro}

Studies of Cosmic Microwave Background (CMB) anisotropy are a top
scientific priority since they address the deepest questions about
origin, structure, and equation of state of the Universe. Given the
sensitivity of current detectors, the main constraint on our ability
to accurately map CMB anisotropies is set by foreground
emissions. While these signals have contaminated, but not dominated,
temperature maps, they are a far bigger problem for CMB
polarization. Also, fighting this contamination is more difficult
because we know much less about polarization than we do about total
intensity emission. The Task Force on CMB Research \citep{boc06}
indeed regards a better characterization of polarized foregrounds as
``a key milestone" in their proposed roadmap.

Extragalactic radio sources are the main CMB contaminant on angular
scales below $0.5^\circ$ at frequencies of up to $\simeq 100$ GHz
\citep{tuc05}. Observational studies of high-frequency polarization
are still scanty and mostly dealing with sources selected at lower
frequencies as shown in \cite{ric04} and in the references in Table 3
of \cite{tuc04}. The blind Australia Telescope 20 GHz (AT20G) survey
of the Southern sky includes polarization measurements at 20, 8.6 and
4.8 GHz; data have been published for the bright source sample
\citep{mas08,bur09}.

The WMAP survey has yielded the first all-sky total intensity and
polarization surveys at frequencies from 23 to 94 GHz. The analysis of
5-yr data \citep{wri09} showed that, in general, the WMAP detected
point sources are not strongly polarized.  Only 5 (Fornax A, Pictor A,
3C 273, Virgo A, and 3C 279) were found to have polarization degrees
greater than 4\% in two or more bands. In this paper we plan to
complement and improve on their analysis in two basic respects. On one
side, we apply a non-blind approach to source detection in
polarization, exploiting the knowledge of positions of the brightest
sources in total intensity, and a new detection technique, called the
filtered fusion technique \citep{arg09}, taking into account the real
beam profiles. On the other side, we check the reliability of our
estimates of polarized flux densities by comparison with the very high
signal-to-noise AT20G BSS (Massardi et al. 2008), \cite{ric04}, and
high frequency VLA calibrator measurements, and by means of numerical
simulations. We also present an estimate of the mean polarization of
sources in a total flux density limited sample.

The polarization is measured by the Stokes parameters $\hat{Q}$,
$\hat{U}$ and $\hat{V}$, and the polarized intensity is
$\hat{P}=\sqrt{ \hat{Q}^2 + \hat{U}^2 + \hat{V}^2}$, see \cite{kam97}
for further details. If we consider linear polarization, $\hat{V}=0$,
which is justified by the fact that extragalactic radio sources have
very low levels of circular polarisation \citep{hom01}, we have to
combine $\hat{Q}$ and $\hat{U}$ maps in an appropriate way to avoid
biasing the estimates of $\hat{P}$. Furthermore, the polarized signal
is just a small fraction of the total intensity signal, of the order
of a few percent, which makes it hard to detect. In the case of WMAP
polarization maps, only a couple of sources can be seen by eye. This
situation may exacerbate the problems with the Eddington (1913) bias
(sources are more easily detected if they happen to lie on top of
positive fluctuations, so that, on average, their fluxes are
overestimated), highlighted by \cite{mas09} for WMAP sources close to
the detection limit in total flux. This issue will be further
investigated.

The outline of this paper is as follows. In \S\,\ref{sec:method} we
describe our method and the sample of WMAP sources, bright in total
flux, to which it was applied to estimate their polarized flux. In
\S\,\ref{sec:results} we discuss the results, also in comparison with
high signal-to-noise ground based measurements at similar
frequencies. Finally, in \S\,\ref{sec:Conclusions}, we summarize our
main conclusions.

\section{Methodology}   \label{sec:method}

\subsection{The Input Catalog} \label{sec:nba}

We have carried out a systematic investigation of the polarized flux
of the 516 sources detected at $\ge 5\sigma$ by \citet{mas09} in the
5-yr WMAP temperature maps, and listed in the NEWPS\_5yr\_5s
catalogue\footnote{http://max.ifca.unican.es/caniego/NEWPS}. 484 of
these sources have a clear identification in low frequency catalogues
(including 27 Galactic objects); the 5 objects detected in
polarization by \cite{wri09} belong to this group. The remaining 32
candidate sources do not have plausible counterparts in all-sky low
radio frequency surveys and may therefore be just high peaks in the
highly non-Gaussian distribution of the other components present in
the maps. If they are all spurious, the reliability of the sample is
94\% and its completeness is of 91\% above 1 Jy.

Since one of our goals is to define a sample as large as possible of
potential calibrators for CMB polarization experiments, we have added
3 extended sources (Cygnus A, Taurus A, Cas A), not included in the
catalog because they lie in very noisy regions, close to the Galactic
equator, but known to be very bright and significantly polarized. The
full sample (Input Catalog) is thus made of 519 sources.

\begin{figure*}
\begin{center}
\includegraphics[width=0.4\textwidth]{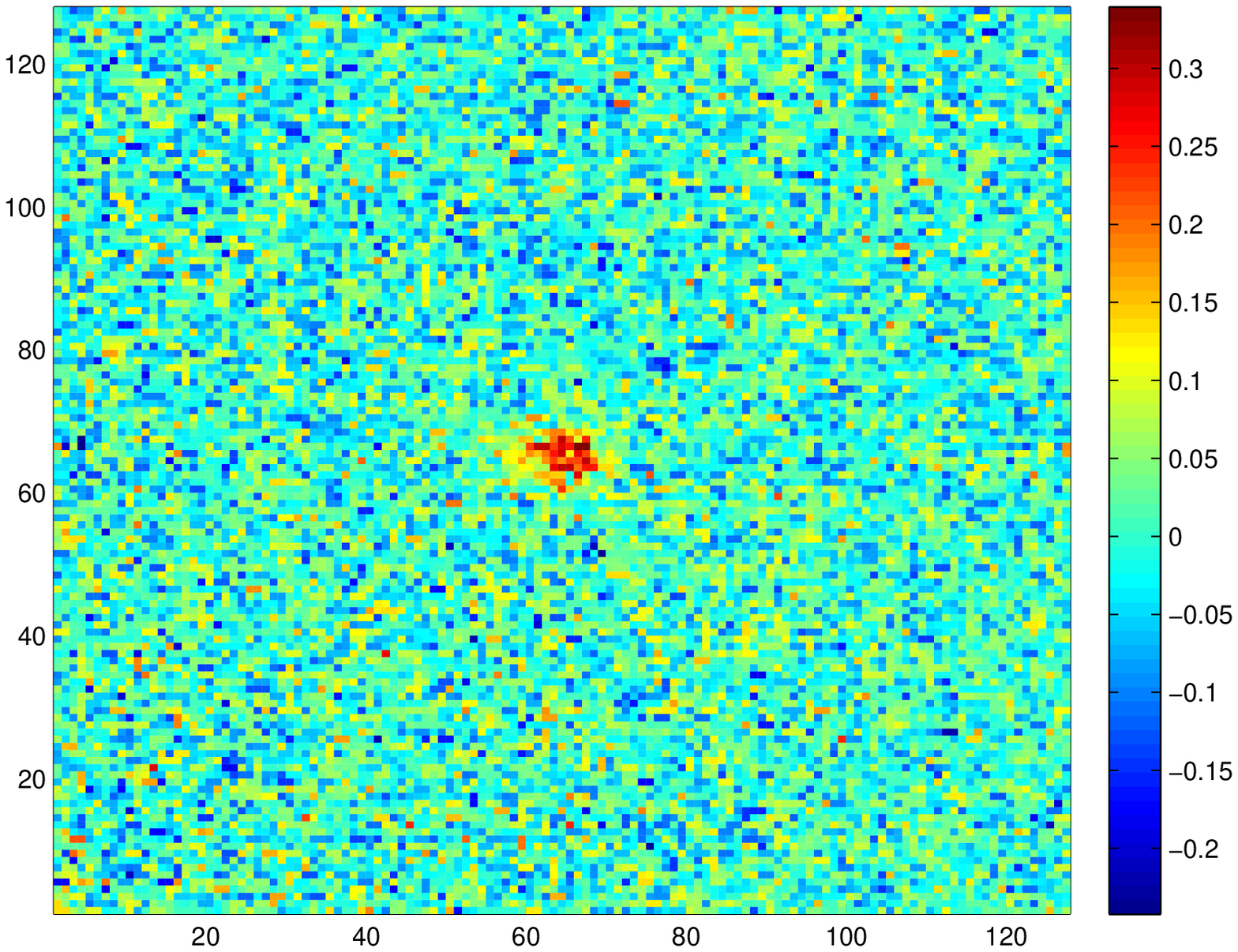}
\includegraphics[width=0.4\textwidth]{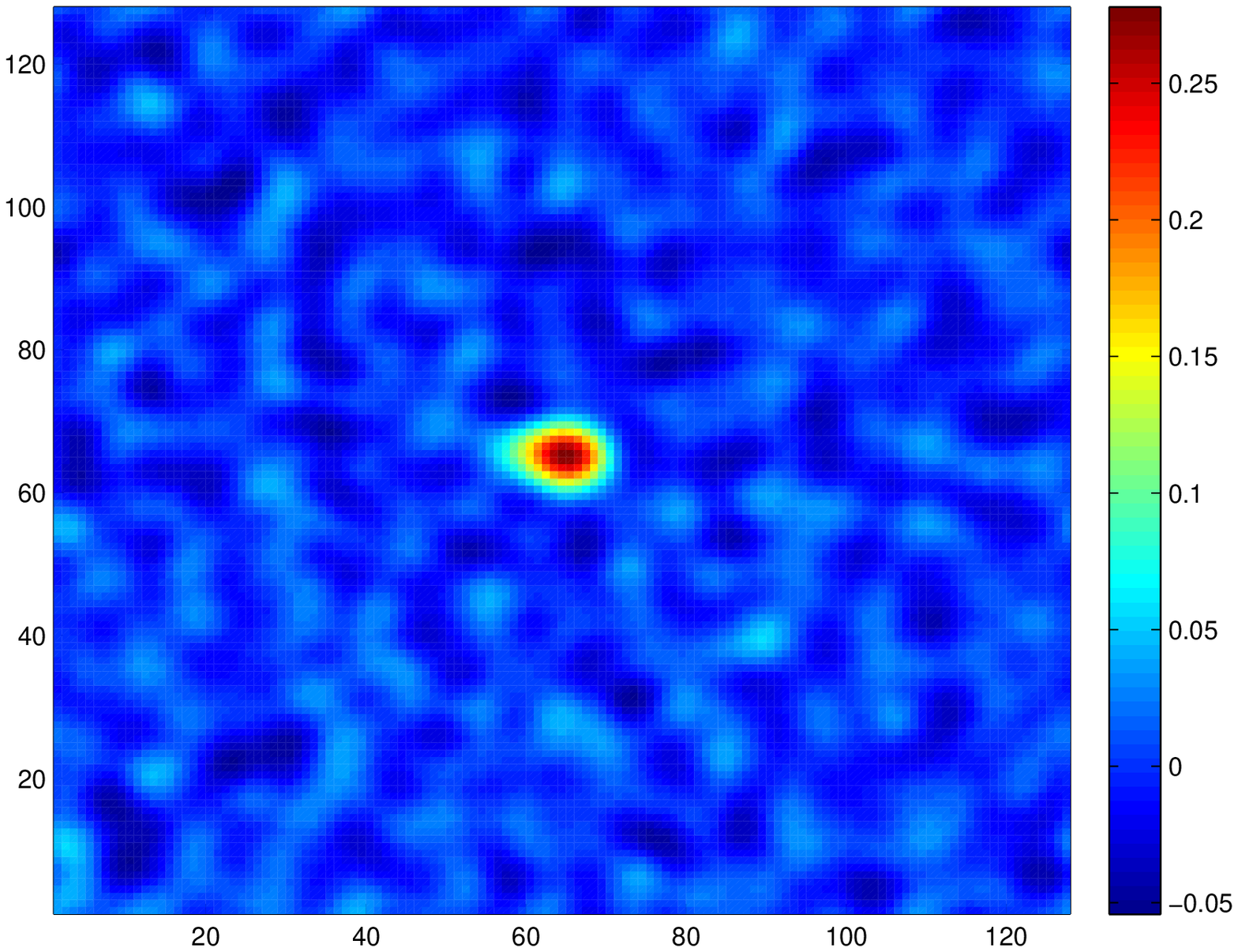}
\includegraphics[width=0.4\textwidth]{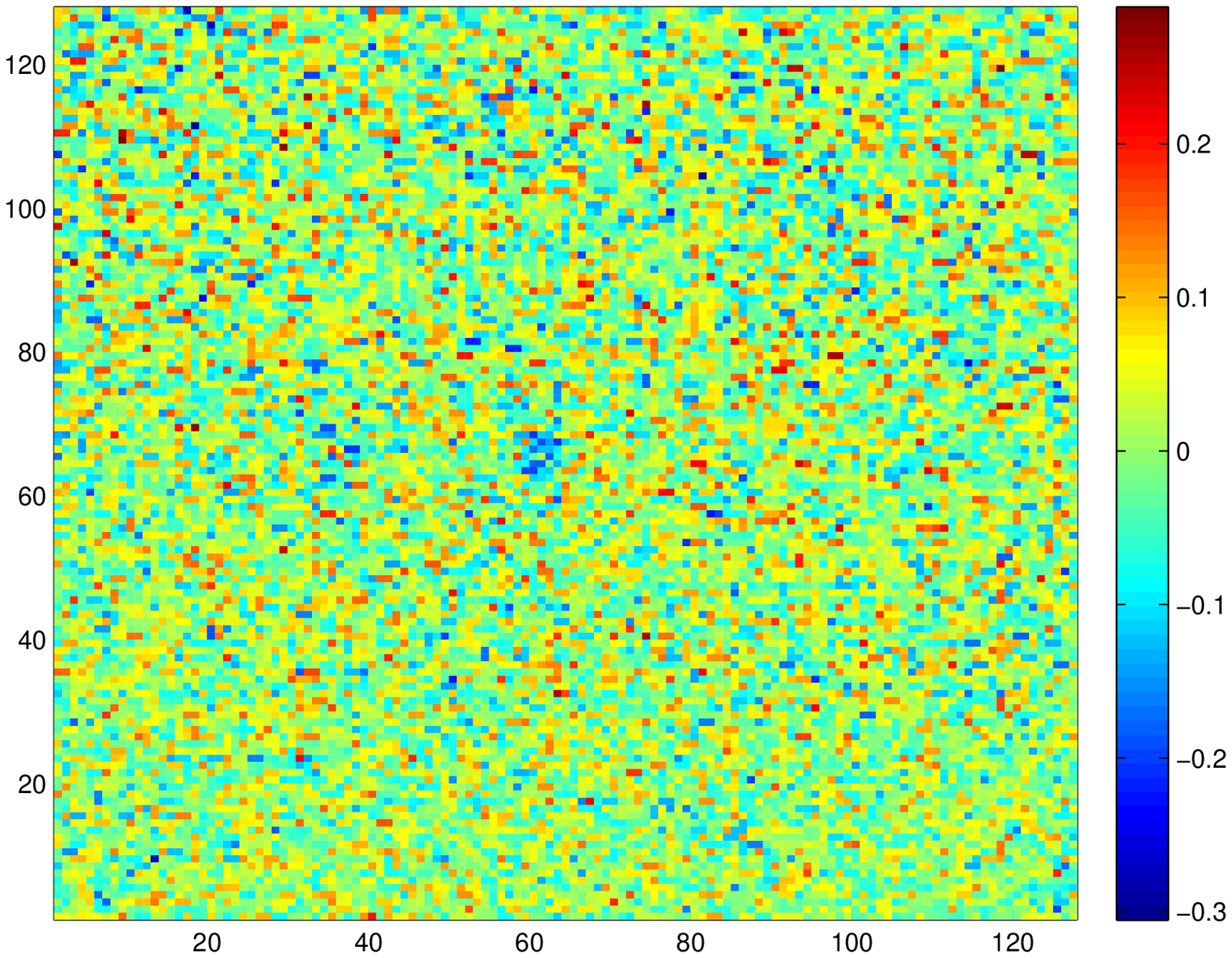}
\includegraphics[width=0.4\textwidth]{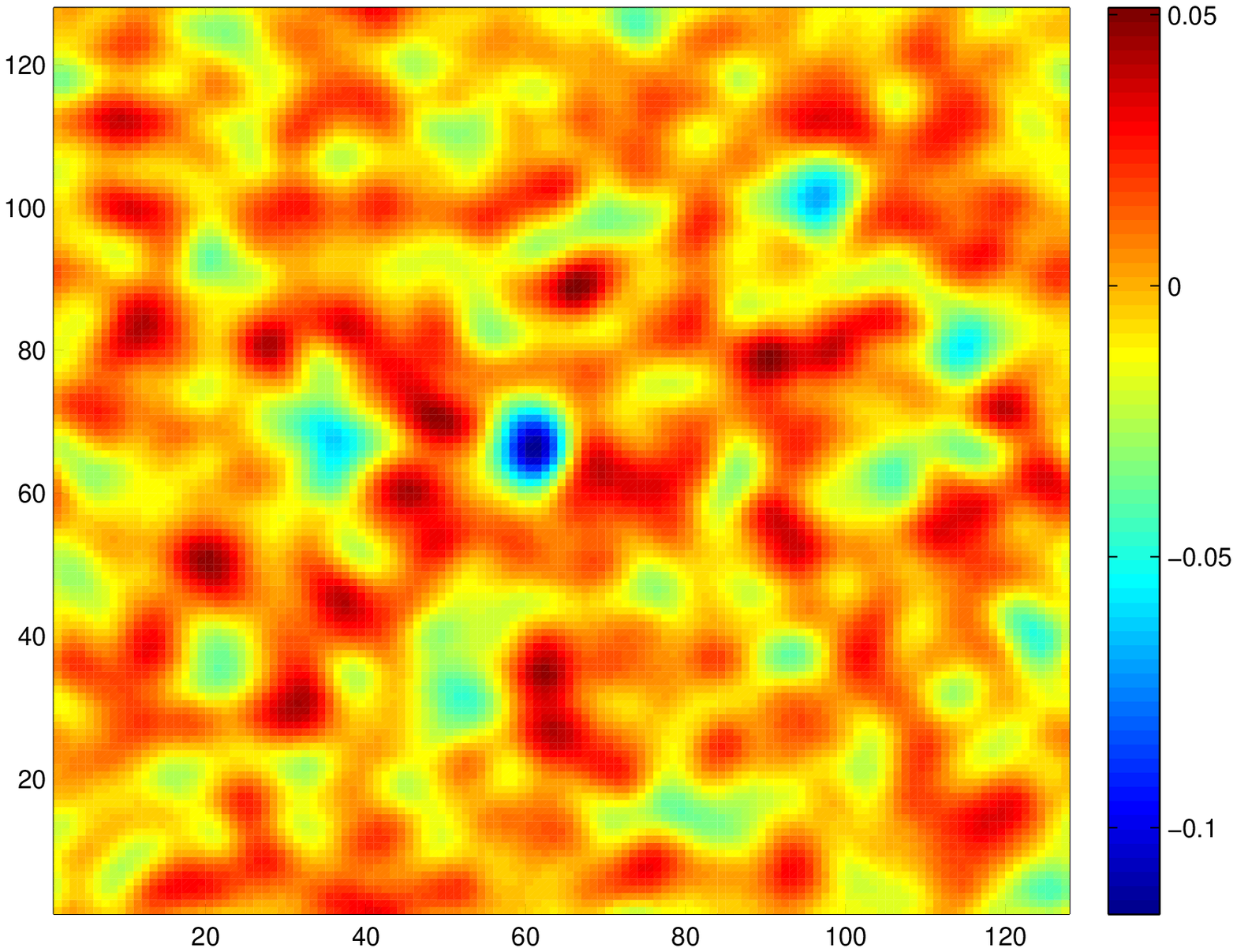}
\includegraphics[width=0.4\textwidth]{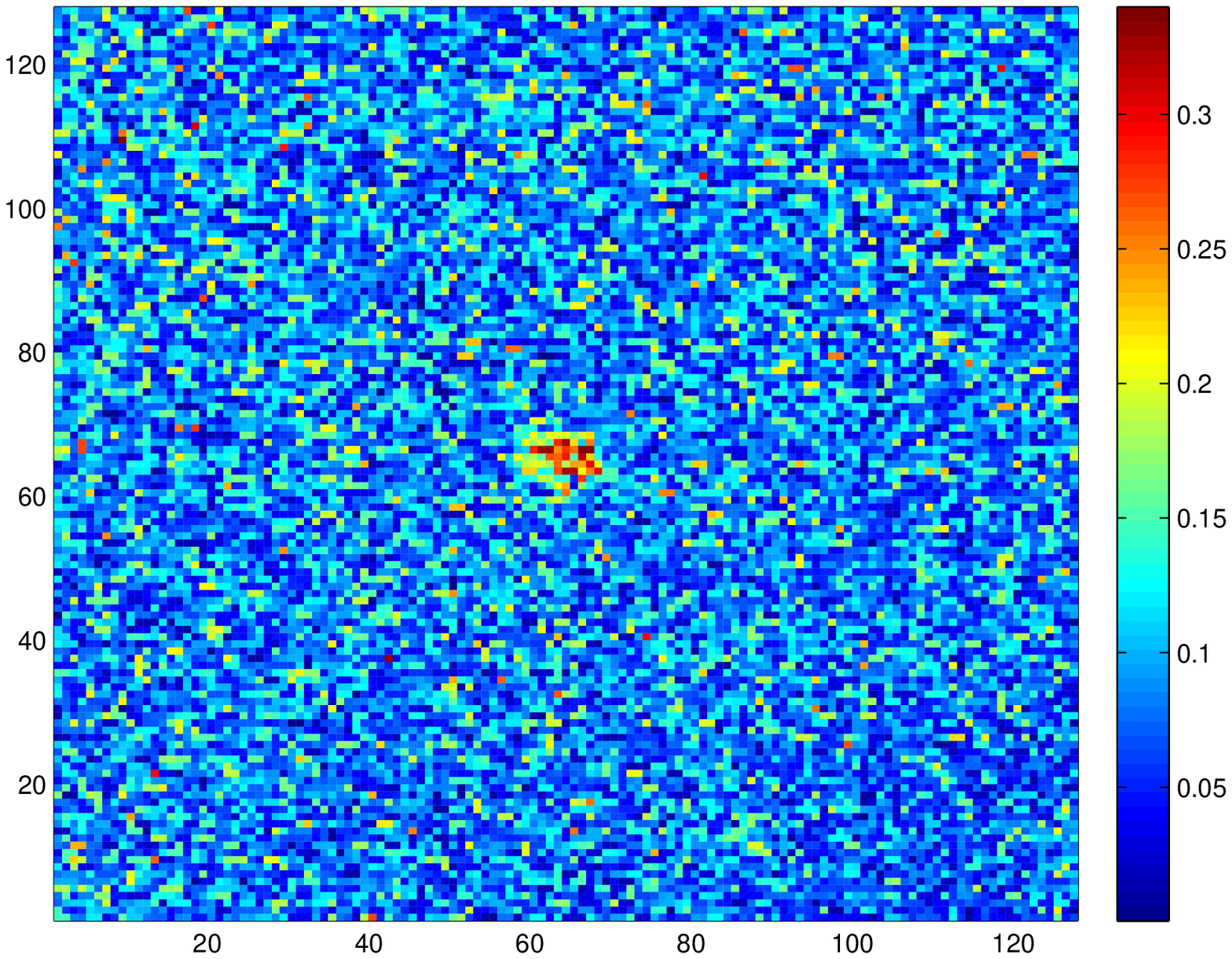}
\includegraphics[width=0.4\textwidth]{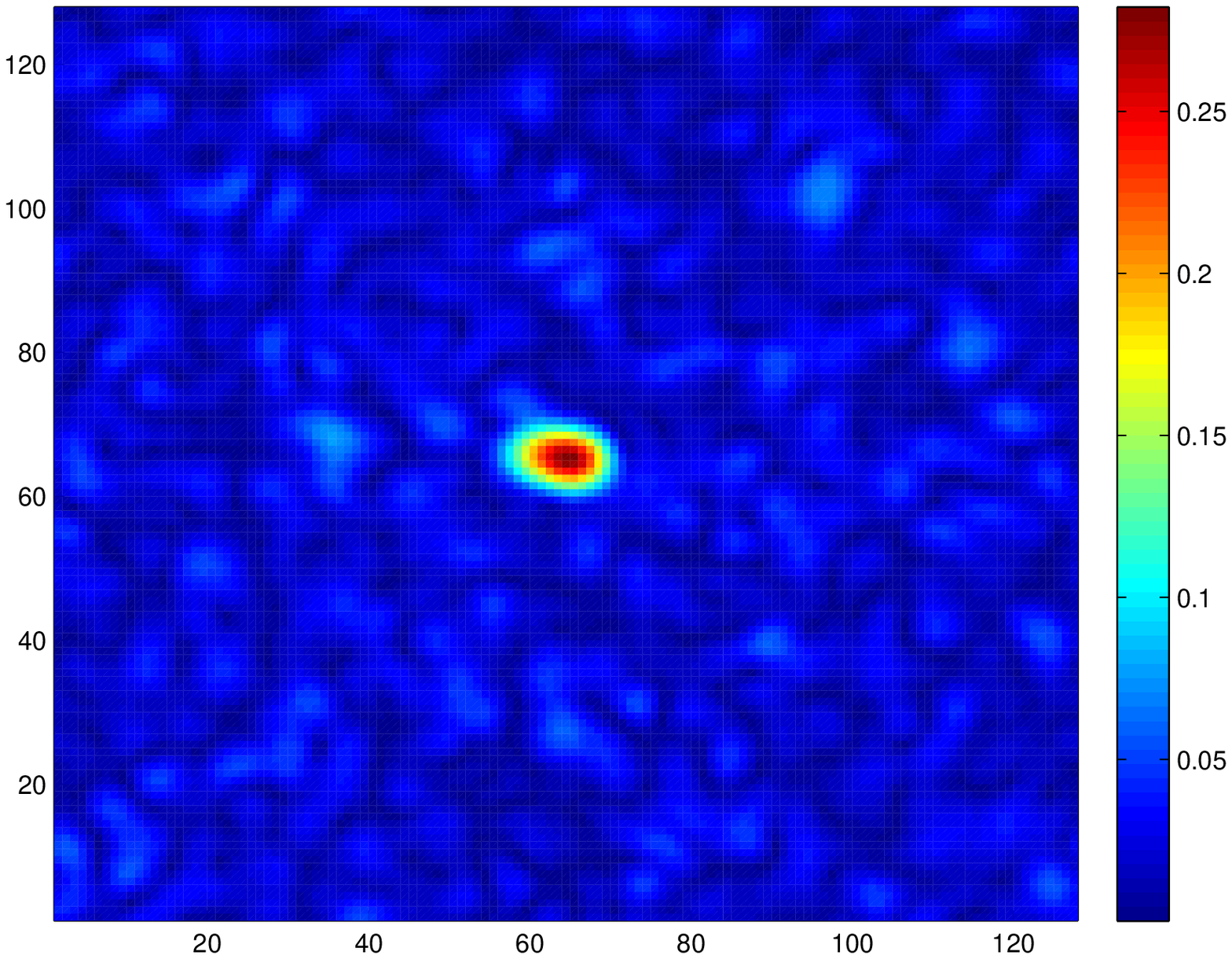}
\caption{Polarization maps at 23 GHz centered on the position of
  Fornax A (no. 74 in the NEWPS\_5yr\_5s catalogue). The upper
    left- and right-hand panels show, respectively, the maps of
    $\hat{Q}$ and ${Q}$ before and after filtering with the matched
    filter, while the middle panels show the analogous $\hat{U}$ and
    ${U}$ maps. The lower panels show, for the same source, the maps
    of $\hat{P}=\sqrt{\hat{Q}^2 + \hat{U}^2}$ (left) and
    $P=\sqrt{Q^2 + U^2}$ (right).} \label{fig:compa}
\end{center}
\end{figure*}

\subsection{Filtered Fusion}

Methods to extract astrophysical foregrounds from multi-frequency CMB
maps frequently exploit the prior knowledge of their frequency
dependence. This approach however does not work well for radio sources
because of the broad variety of their spectral properties in the
relevant frequency range \citep{mas08,sad08}. On the other hand, with
few exceptions, extragalactic radio sources look point-like when
observed with the beams used by CMB experiments, and therefore have,
in the maps, the shape of the effective angular response function of
the instrument. In the literature one can find several methods
exploiting this property and using linear filters to detect point
sources. The standard matched filter approach has been used for years
\citep{nai92,vik95,mal97,teg98,san01,her02,ste06}. More recently, a
multi-frequency approach based on matched filters has been elaborated
\citep{her08a,her08b}. An approach based on optimal wavelets
\citep{vie01,vie03,bar04,san06,gon06} was successfully applied to WMAP
maps \citep{lop07,mas09} as well as to realistic simulations of Planck
maps \citep{lop06,lea08}.  Moreover, filters based on the
Neyman-Pearson approach, using the distribution of maxima, have been
proposed (L{\'o}pez-Caniego et al. 2005a,b) and a Bayesian approach
has been developed \citep{hob03,fer08,car09}.

In this work, following Arg\"ueso et al. (2009), we use the same matched filter
over $Q$ and $U$ images. The matched filter is a circularly-symmetric filter,
$\Psi(x;R,b)$, such that the filtered map, $w(R, b)$ satisfies the following
two conditions: $(1) \ \langle w(R_0, 0)\rangle = s(0) \equiv A$, i.e. $w(R,
0)$ is an \emph{unbiased} estimator of the flux density of the source; $(2)$
the variance of $w(R, b)$ has a minimum on the scale $R_0$, i.e. it is an
\emph{efficient} estimator. In Fourier space the matched filter writes as:
\begin{equation} \label{eq:mf}
  \psi_{MF} = \frac{1}{a}\frac{\tau (q)}{\Delta(q)}, \ \ \
  a = 2 \pi \int dq q \frac{\tau^2 (q)}{\Delta(q)},
\end{equation}
\noindent where $\Delta(q)$ is the power spectrum of the background
and $\tau (q)$ is Fourier transform of the source profile (equal to
the beam profile for point sources).

Since, in this application, each patch is centered on the position of
a source detected in total intensity, we describe the source as
\begin{equation} \label{eq:source}
s(\vec{x})=A\tau(\vec{x}),
\end{equation}
\noindent where $A$ is its unknown polarized flux density and
$\tau(\vec{x})$ is its \emph{profile}. We assume circular symmetry, so
that $\tau(\vec{x})=\tau(x)$, $x=|\vec{x}|$. For point sources the
profile is equal to the beam response function of the detector. The
WMAP beams are not Gaussian and we use the real symmetrized radial
beam profiles for the different WMAP channels to construct our
filters.

The matched filter gives directly the maximum amplification of the
source and yields the best linear estimation of the flux. As
extensively discussed in the literature, it is a very powerful tool to
detect point sources, but it has to be used with care because its
performance may degrade rapidly if it is not properly implemented. In
particular, the power spectrum $\Delta(q)$ of the image has to be
obtained directly from the data.

The WMAP team has used a matched filter that operates on the sphere
and takes into account the non-Gaussian profile of the beam. It is
described by eq.~(\ref{eq:mf}), replacing the flat limit quantities
$\tau(q)$ and $\Delta(q)$ with their harmonic equivalents $b_{\ell}$
and $C_{\ell}$. As pointed out by \cite{lop06}, the use of the
$C_\ell$'s computed from all-sky maps to construct a global matched
filter that operates on the sphere is a reasonable first approach, but
we think that it can be improved by operating locally.

We have followed the scheme named {\it filtered fusion} by its authors
\citep{arg09}. For each object in the input catalog, and for each WMAP
frequency between 23 and 61 GHz, we have projected two patches (one
for $Q$ and one for $U$), each of $14.65\times 14.65$ square degrees,
centered on the source position. We have left aside the 94 GHz channel
because of the normalization problems discussed by L{\'o}pez-Caniego
et al. (2007) and Gonz{\'a}lez-Nuevo et al. (2008).  Each patch is
made of $128\times 128$ 6.87 arcmin pixels (HealPix
$\hbox{Nside}=512$, \cite{gor05}). The projection has been done using
the CPACK
library\footnote{http://astro.ic.ac.uk/mortlock/cpack/}. Then, each
pair of patches has been filtered using a matched filter exploiting
the power spectrum determined within each patch. After filtering, the
$Q^2$ and $U^2$ are added together and the square root of the
resulting image has been calculated. The noise bias can be removed by
subtracting from $Q^2$ and $U^2$ the corresponding noise contribution
$\sigma^2_{\hat{Q}}$ and $\sigma^2_{\hat{U}}$. This correction turns
out to be negligible. In this way we have obtained a map of the
polarized intensity $P$ within each patch. This approach differs from
the usual one, where a $\hat{P}$ map is constructed adding together
the unfiltered $\hat{Q^2}$ and $\hat{U^2}$ maps, and taking the square
root of the resulting map (see Fig.~\ref{fig:compa}).

We have then looked for the brightest maximum inside a circle
centered on the center of each patch and covering the area of the WMAP
beam. Next, we estimated the noise level of the patch and the
significance of the possible detection.  Finally, we constructed
catalogues containing all sources whose polarized flux $P$ was
detected above a chosen significance level.

The reason why we need to use the significance of the detection
instead of the usual signal-to-noise ratio is that the noise does not
obey a Gaussian distribution but a Rayleigh distribution, since we are
dealing with maps that have been squared. The significance was derived
from the distribution of the values of $P$ for the pixels within the
patch. An example of such distribution is in Fig.~\ref{fig:histo}.

\subsection{Flux and error estimation} \label{sec:pfe}

The polarized flux densities and their errors were estimated in a way
similar to that applied for total intensity (Massardi et
al. 2009). Point sources appear in the image with a profile identical
to the beam profile. For example, if the beams were Gaussian, the
source flux could be obtained multiplying the flux in the brightest
pixel by the ratio between the beam and the pixel area,
$2\pi(R_s)^2/L_p^2$, where $R_s=\mathrm{FWHM}/(2\sqrt{2\log{2}})$ and
$L_p$ is the pixel side.

In our case the beams are not Gaussian and we need to calculate this
relationship integrating over the real symmetrized beam profile for
each channel. In doing that we have to take into account that we work
with HEALPix pixelization \citep{gor05} at Nside=512. Although the
image is centered at the position of the source, after the projection
to the flat patch the object does not always lie in the central pixel,
but may end up in an adjacent one. Thus, to estimate its flux we make
reference not to the intensity in the central pixel but to that of the
brightest pixel close to the center of the \emph{filtered} image
within an area equal to that of the beam. As discussed in \S\,3.3 of
\cite{lop07}, this method for flux estimation through linear filtering
is almost optimal in many circumstances.

\begin{figure}
\begin{center}
\includegraphics[width=0.5\textwidth]{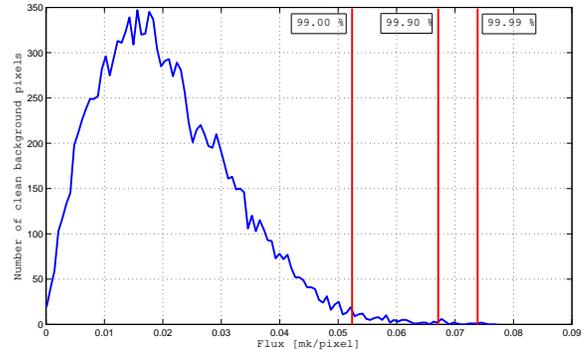}
\caption{Histogram showing the distribution of the polarized flux
  $P=\sqrt{(Q^2 + U^2)}$ (in mK per pixel) obtained from a filtered
  WMAP 23 GHz patch of $\hat{Q}$ and $\hat{U}$ centered on the
  position of Fornax A (14.65 square degree patch and 128x128
  pixels). This histogram has been produced with the values of $\sim
  13500$ pixels, excluding the flagged ones (see
  Fig.~\protect\ref{fig:mask}). The vertical lines correspond to the
  values of $P$ exceeding those of $99.0\%$, $99.9\%$, and $99.99\%$
  of the pixels. In other words, measured values of $P$ at these
  levels have, respectively, $99.0\%$, $99.9\%$, and $99.99\%$
  probability of {\it not} being due to noise spikes. Note that the
  polarized flux of a source is obtained multiplying the flux in the
  brightest pixel by the ratio between the beam and the pixel area
  (see \S\,\protect\ref{sec:pfe}).}
\label{fig:histo}
\end{center}
\end{figure}

\begin{figure}
\begin{center}
\includegraphics[width=0.45\textwidth]{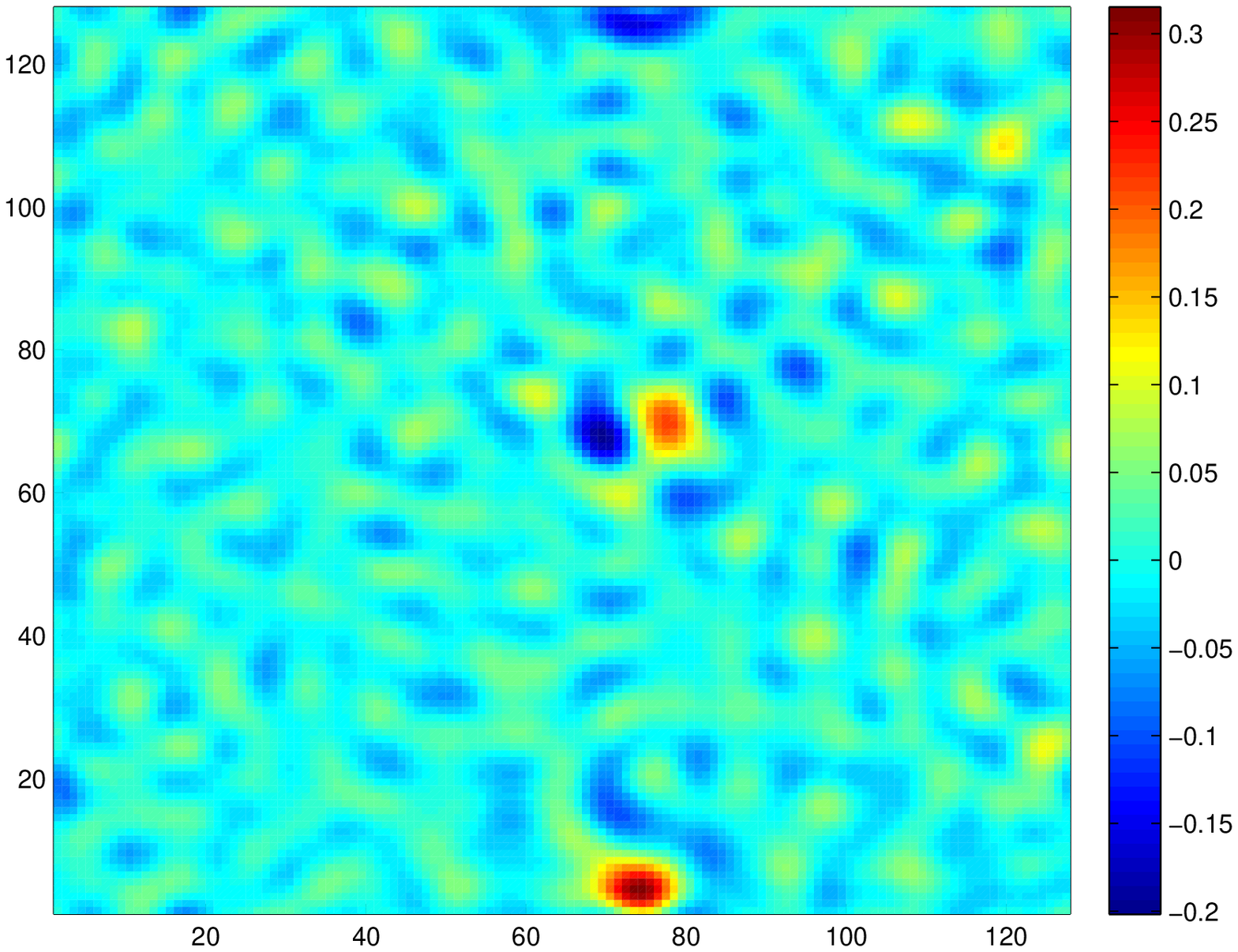}
\includegraphics[width=0.45\textwidth]{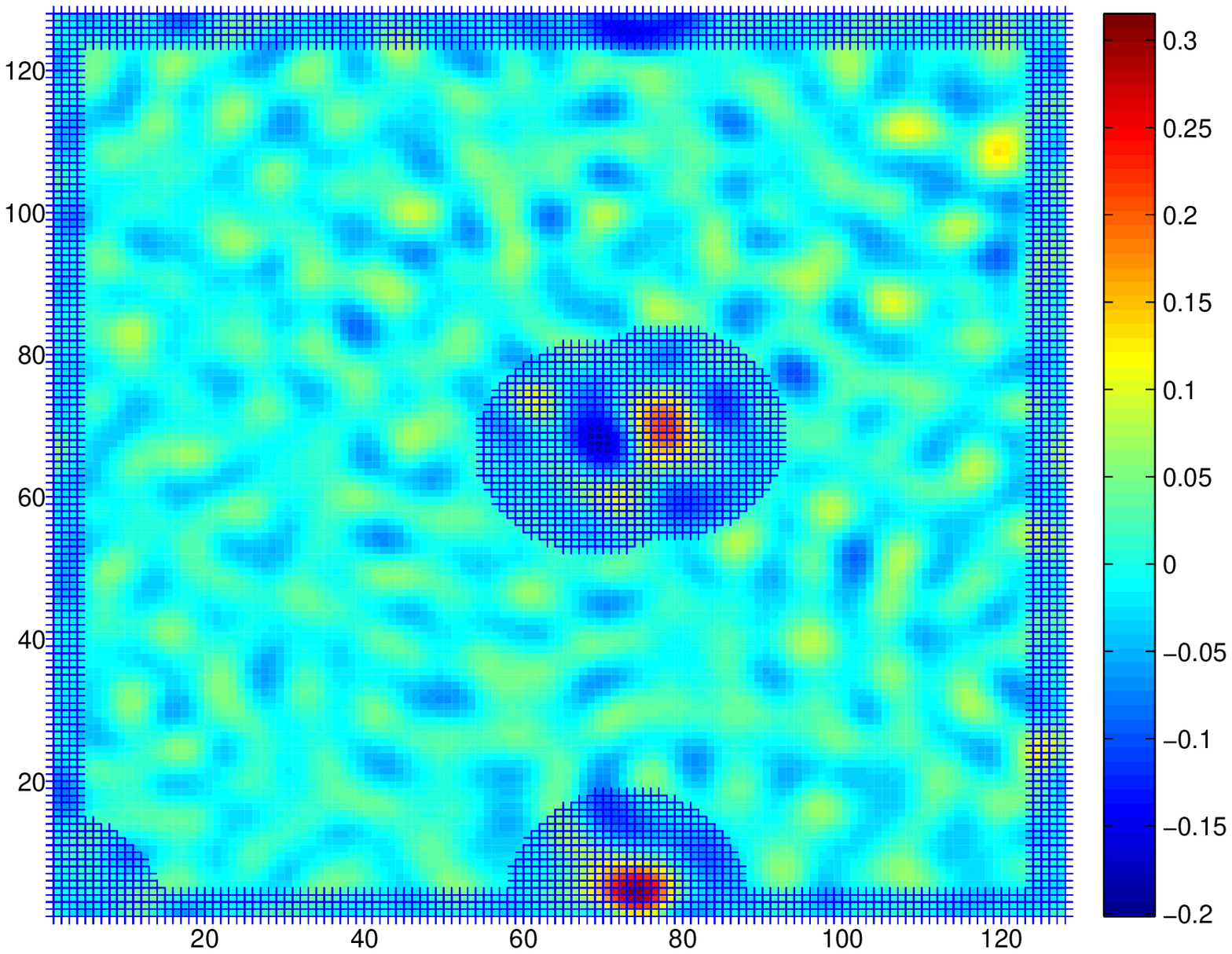}
\caption{These panels show an example of the flagging of the borders and the brightest $5\%$ objects
in a given patch to prevent contaminated pixels to be used in the calculation of the
background distribution. The upper panel shows a filtered patch where the bright
sources are identified. The lower panel shows, cross-hatched, the flagged area.
\label{fig:mask}}
\end{center}
\end{figure}

Note that the method adopted here to estimate the flux differs from
the one used by the WMAP team. Assuming that they have followed
similar procedures in intensity and polarization, they used a matched
filter taking into account the non-Gaussian profile of the beam to
detect point sources in the filtered full-sky maps, but their fluxes
have been derived fitting the pixel intensities around the point
source to a Gaussian profile plus a plane baseline (in the unfiltered
image).

We calculated the rms noise for each patch containing an input
source. This value can be easily overestimated if border effects and
strong fluctuations due to other point sources or small scale
structure of the diffuse emissions in the patch are not removed or
filtered out. In order to avoid this, first we find the 5\% brightest
pixels in the patch and flag them. Second, we flag pixels within a
distance equal to 15 pixels from the border (see
Fig.~\ref{fig:mask}). Finally, we select a shell around the source
with inner radius equal to the FWHM of the beam (since polarized
fluxes are never very high, this is enough for them to have decreased
well below the noise level), and outer radius of $3\times\hbox{FWHM}$.
The rms noise, $\sigma$, is then obtained as the square root of the
variance of the pixels included in this shell, excluding flagged
pixels (if any). From the distribution of the values of $P$ in the
unflagged pixels we calculate the levels exceeding those of 95.00\%,
99.00\%, 99.90\%, and 99.99\% of pixels. These are the probabilities
that signals at those levels are real rather than noise
fluctuations. An example is shown in Fig.~\ref{fig:histo}.

\begin{deluxetable}{crrrr}
\tabletypesize{\scriptsize} \tablewidth{0pt} \tablecaption{Number of detections at different significance levels\label{tab:dets}}
 \tablehead{ \textbf{Significance} & \textbf{23\ GHz} & \textbf{33\ GHz} & \textbf{41\ GHz} & \textbf{61\ GHz}  }
    \startdata
    \textbf{95.00\%} & 138 & 122 & 93 & 81  \\
    \textbf{99.00\%} &  53 & 41 & 30 & 20 \\
    \textbf{99.90\%} &  20 & 16 & 11 & 6 \\
    \textbf{99.99\%} &  18 & 12 & 9  & 6 \\ \hline
\end{deluxetable}

\begin{deluxetable*}{lrrrrrrrrr}

\tabletypesize{\scriptsize} \tablewidth{0pt} \tablecaption{The POlarized WMAP Point Sources (POWPS) sample.
\label{tab:powps}}
\tablehead{ \multicolumn{1}{c}{\textbf{Object}} & \multicolumn{1}{c}{\textbf{$RA$}} & \multicolumn{1}{c}{\textbf{$DEC$}}
& \multicolumn{1}{c}{\textbf{$GLON$}} & \multicolumn{1}{c}{\textbf{$GLAT$}} & \multicolumn{1}{c}{\textbf{$P$} (Jy)}
& \multicolumn{1}{c}{\textbf{$P$} (Jy)} & \multicolumn{1}{c}{\textbf{$P$} (Jy)} & \multicolumn{1}{c}{\textbf{$P$} (Jy)}
& \multicolumn{1}{c}{\textbf{flags}}\\
& \multicolumn{1}{c}{[h]} & \multicolumn{1}{c}{[deg]} & \multicolumn{1}{c}{[deg]} & \multicolumn{1}{c}{[deg]} &
\multicolumn{1}{c}{23\ GHz} & \multicolumn{1}{c}{33\ GHz} & \multicolumn{1}{c}{41\ GHz} &  \multicolumn{1}{c}{61\ GHz}
\\ }

\startdata

\textbf{ \phantom{1}74 } (Fornax A) &   3.372 & -37.177 & 240.122 & -56.766 &   $1.07\pm 0.04$ &   $0.87\pm 0.06$ &   $0.49\pm 0.06$ &   $0.41\pm 0.10$ & 1 1 1 4 \\
\textbf{ 126 } (Pictor A) &   5.326 & -45.743 & 251.548 & -34.680 &  $0.38\pm 0.05$ &   $0.42\pm 0.06$ &   $0.41\pm 0.07$ &   \multicolumn{1}{c}{--} & 1 1 3 0 \\
\textbf{ 156 } (PKS0607-15)&   6.166 & -15.679 & 222.609 & -16.109 & $0.33\pm 0.04$ &   $0.36\pm 0.07$ &   \multicolumn{1}{c}{--} &   \multicolumn{1}{c}{--} & 1 3 0 0 \\
\textbf{ 200 } (PKS0829+04)&   8.530 &   4.559 & 220.627 &  24.357 & \multicolumn{1}{c}{--} & \multicolumn{1}{c}{--} &   $0.56\pm 0.09$ &   $1.10\pm 0.16$ & 0 0 3 1 \\
\textbf{ 256 } (PKS1144-37)&  11.787 & -38.150 & 289.265 &  23.012 & \multicolumn{1}{c}{--} & $0.49\pm 0.07$ &   \multicolumn{1}{c}{--} &   \multicolumn{1}{c}{--} & 0 1 0 0 \\
\textbf{ 266 } (NC)&  12.200 & -52.630 & 296.882 &   9.778 &   $0.73\pm 0.04$ &   $0.42\pm 0.06$ &   $0.44\pm 0.07$ & \multicolumn{1}{c}{--} & 1 1 1 0 \\
\textbf{ 272 } (3C273)&  12.485 &   2.044 & 289.957 &  64.352 & $1.07\pm 0.05$ &   $0.71\pm 0.06$ &   $0.72\pm 0.08$ &   \multicolumn{1}{c}{--} & 1 1 1 0 \\
\textbf{ 273/274 } (Virgo A)&  12.509 &  12.350 & 283.597 &  74.433 &   $0.79\pm 0.05$ &   $0.71\pm 0.07$ & $0.62\pm 0.08$  & $0.53\pm 0.14$ & 1 1 1 1 \\
\textbf{ 280 } (3C279)&  12.936 &  -5.762 & 305.107 &  57.090 &   $0.67\pm 0.05$ &   $0.45\pm 0.08$ &   $0.83\pm 0.09$ & $0.66\pm 0.17$ &   1 3 1 4 \\
\textbf{ 289 } (PKS1320-44)&  13.381 & -44.682 & 308.790 &  17.832 &   $1.67\pm 0.08$ &   $1.13\pm 0.12$ &   $0.56\pm 0.13$ &   \multicolumn{1}{c}{--} & 1 1 3 0 \\
\textbf{ 291 } (Centaurus A)&  13.422 & -43.025 & 309.483 &  19.416 & $3.19\pm 0.08$ &   $2.30\pm 0.11$ &   $2.02\pm 0.11$ &   $1.58\pm 0.13$ & 1 1 1 1 \\
\textbf{ 295 } (3C286)&  13.527 &  30.510 &  56.332 &  80.578 & $0.33\pm 0.04$ &   $0.50\pm 0.06$ &   \multicolumn{1}{c}{--} &   $0.63\pm0.11$ & 1 1 0 3 \\
\textbf{ 337 } (PKS1546+02)&  15.823 &   2.545 &  10.743 &  40.891 & $0.33\pm 0.05$ &   \multicolumn{1}{c}{--} &   \multicolumn{1}{c}{--} &   \multicolumn{1}{c}{--} &  1 0 0 0 \\
\textbf{ 346 } (NC)&  16.345 & -25.487 & 351.319 &  17.150 & $0.45\pm 0.06$  &   \multicolumn{1}{c}{--} &   \multicolumn{1}{c}{--} &   \multicolumn{1}{c}{--} & 1 0 0 0 \\
\textbf{ 432 } (NC)&  20.278 &  45.776 &  82.152 &   5.810 & $1.16\pm 0.07$  &   $0.67\pm 0.07$ & $0.38\pm 0.07$ & \multicolumn{1}{c}{--} & 1 1 3 0 \\
\textbf{ 437 } (GB6 J2038+51)&  20.647 &  51.316 &  88.822 &   6.016 & $0.62\pm 0.07$ &   $0.29\pm 0.07$ &   $0.40\pm 0.07$ &   \multicolumn{1}{c}{--} & 1 4 3 0 \\
\textbf{ 439 } (NC)&  20.846 &  29.160 &  72.753 &  -9.460 &   $0.45\pm 0.05$ &   $0.33\pm 0.06$ &   $0.31\pm 0.07$ &   $0.48\pm 0.11$ & 1 3 4 4 \\
\textbf{ 440 } (G93.3+6.9)&  20.871 &  55.404 &  93.307 &   6.956 & $0.57\pm 0.06$ &   $0.36\pm 0.07$ &   $0.33\pm 0.06$ &   \multicolumn{1}{c}{--} & 1 4 4 0 \\
\textbf{ 473 } (NC)&  22.323 &  26.438 &  85.400 & -25.138 & \multicolumn{1}{c}{--} &   \multicolumn{1}{c}{--} &   \multicolumn{1}{c}{--} &   $0.95\pm 0.13$ & 0 0 0 1 \\
\textbf{ Cygnus A } (3C405)&  19.984 &  40.484 &  75.930 &   5.700 & $0.49\pm 0.07$ & $0.51\pm 0.07$ &   $0.56\pm 0.06$ &   $0.48\pm 0.12$ &  3 2 1 4 \\
\textbf{ Taurus A } (Crab)&   5.583 &  22.369 & 184.310 &  -5.510 & $24.7\pm 0.18$ &  $20.2\pm 0.16$ &  $16.0\pm 0.14$ &  $6.41\pm 0.18$ & 1 1 1 1 \\
\textbf{ Cas A } (3C461) & 23.408 & 58.835 & 111.872 & -2.157 &
$0.91\pm 0.05$ & $0.55\pm 0.06$ & $0.58\pm 0.06$ &
$0.22\pm 0.11$ & 1 1 1 1 \\
\tablecomments{Sources detected at $>99.99\%$ confidence level in at
  least one of the WMAP frequency channels. Column 1: sequential
  number in the NEWPS\_5yr\_5s catalogue and source name (NC means
  that the source has no plausible low radio frequency counterpart;
  see text); columns 2-5: equatorial (J2000) and Galactic coordinates
  of the source; columns 6-9: detected integrated polarized flux
  density and their errors at 23, 33, 41 and 61 GHz; columns 10-13:
  flags for significance ($1:~\ge 99.99\%$, $2:~\ge 99.90\%$ but $<
  99.99\%$, $3:~\ge 99.00\%$ but $< 99.90\%$, $4:~\ge 95.00\%$ but $<
  99.00\%$, $0= < 95.00\%$), at 23, 33, 41 and 61 GHz,
  respectively. The NEWPS\_5yr\_5s catalogue lists 2 sources (No. 273
  and 274) close to the position of Virgo A. The present re-analysis
  has shown that they are actually the same source, coinciding with
  Virgo A with total flux density of $18.4\pm 0.25$ at 23 GHz. }
\end{deluxetable*}

\begin{deluxetable*}{ccccccc}
\tabletypesize{\scriptsize} \tablewidth{0pt} \tablecaption{Comparison of polarized flux estimates in the present paper
with those of \cite{wri09} \label{tab:wmap}}
 \tablehead{ \textbf{Object} & \textbf{$23\,GHz_{\rm W}$}
 & \textbf{$23\,GHz_{\rm p}$} & \textbf{$33\,GHz_{\rm W}$} & \textbf{$33\,GHz_{\rm p}$} &
 \textbf{$41\,GHz_{\rm W}$} & \textbf{$41\,GHz_{\rm p}$} }
    \startdata
    \textbf{0322-3711} (For A) & $1.57\pm 0.07$ & $1.07\pm 0.04$ & $1.17\pm 0.15$ & $0.87\pm 0.06$ & $0.85\pm 0.23$ & $0.49\pm 0.06$ \\
    \textbf{0519-4546} (Pic A) & $0.39\pm 0.06$ & $0.38\pm 0.05$ & $0.45\pm 0.10$ & $0.42\pm 0.06$ & -- & $0.41\pm 0.06$ \\
    \textbf{1229+0203} (3C273) & $0.98\pm 0.06$ & $1.07\pm 0.05$ & $0.81\pm 0.11$ & $0.71\pm 0.06$ & $0.80\pm 0.13$ & $0.72\pm 0.08$ \\
    \textbf{1230+1223} (Vir A) & $0.75\pm 0.08$ & $0.79\pm 0.05$ & $0.74\pm 0.11$ & $0.71\pm 0.07$ & $0.50\pm 0.11$ & $0.62\pm 0.08$ \\
    \textbf{1256-0547} (3C279) & $0.62\pm 0.07$ & $0.67\pm 0.05$ & $0.55\pm 0.12$ & $0.45\pm 0.08$ & $0.69\pm 0.15$ & $0.83\pm 0.09$ \\

    \tablecomments{The subscripts `W' and `p' designate the columns containing the results by \cite{wri09} and of the
    present paper, respectively. }
\end{deluxetable*}

\begin{figure*}
\begin{center}
\includegraphics[width=0.6\textwidth, angle=90]{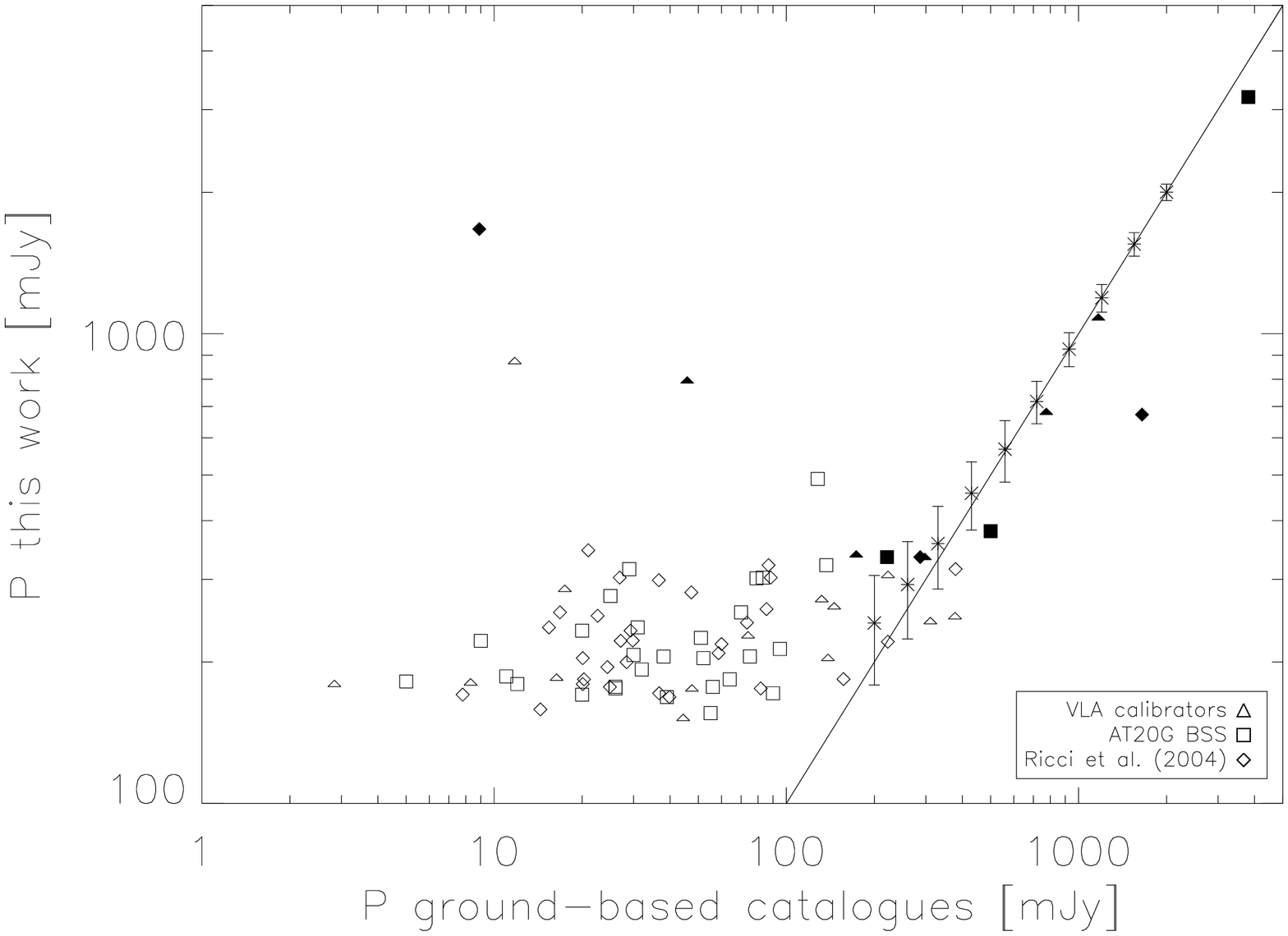}
\caption{Comparison of polarized flux density estimates from the WMAP
  23 GHz map (present paper) and ground-based observations: AT20G BSS
  (Massardi et al. 2008; squares); Ricci et al. (2004) (diamonds); VLA
  calibrators (triangles). Filled symbols correspond to sources
  detected on the WMAP map with more than 99.99\% significance
  level. In case of sources with multiple ground-based observations,
  we have chosen those with the resolution closest to WMAP's. The
  solid line corresponds to equal values on the two axes. The two
  highest filled symbols on the left of the solid line are examples of
  the effect of the much higher resolution of ground based
  measurements, compared to WMAP's; the filled diamond is Cen A
  (resolved also by WMAP), and the filled triangle is Virgo A
  (unresolved by WMAP). The highest open triangle corresponds to a
  source close to the Galactic plane, whose polarized flux estimated
  from the WMAP maps includes a dominant Galactic contribution. The
  outlier on the right of the solid line is the highly variable blazar
  3C279. The asterisks with error bars show the results of simulations
  described in\S\,\protect\ref{sec:sim}\label{fig:comparisons}}
\end{center}
\end{figure*}

\section{Results} \label{sec:results}

In Table~\ref{tab:dets} we list the number of objects with $P$ values
significant at more than 95.00\%, 99.00\%, 99.90\%, and 99.99\%
confidence levels. There are 18, 12, 9, and 6 detections with $\ge
99.99\%$ significance level at 23, 33, 41, and 61 GHz,
respectively. The 22 sources detected at such significance level at,
at least, one frequency, are listed in Table \ref{tab:powps}. Several
of them are well known bright extended objects including Cassiopeia A,
Centaurus A, Cygnus A, Fornax A, Pictor A, Taurus A (Crab Nebula), and
Virgo A (but, except for Cen A and For A, their sizes are smaller than
the WMAP 23 GHz beam). Since our algorithm is not optimized for
extended sources, the flux estimates for these objects are likely
affected by systematic errors much larger than the quoted statistical
errors. In addition resolution effects can be seen in WMAP data, for
example they likely account for the steep drop of the Crab polarized
flux between 41 and 61 GHz. Such drop corresponds to the ratio of WMAP
beam areas at these two frequencies. Five objects do not have
plausible counterparts in low frequency radio surveys. They may
therefore be exceptionally high peaks in the highly non-Gaussian
fluctuation field, in temperature and polarization, mostly due to
Galactic synchrotron (with possible CMB and source confusion
contributions). In fact, 3 of these 5 objects are at $|b|< 10^\circ$
and a fourth one (no. 346) is in the Ophiuchus Complex region.

All the 5 objects detected in polarization by \cite{wri09} have $P$
values significant at $>99.99\%$ levels at least at 1 frequency. As
shown by Table~\ref{tab:wmap} our flux estimates are in generally good
agreement with those by \cite{wri09}, in spite of the
different techniques used. The main formal discrepancy concerns the 23
GHz polarized flux of 0322-3711 (Fornax A), a well known extended
source for which we expect that photometric errors are mostly
systematic, as noted above. 

To better assess the reliability of our estimates we have compared
those sources detected at $\ge 95.00\%$ confidence levels with ground
based measurements at nearby frequencies. Unfortunately there are only
few samples that can be used for such a comparison. The AT20G Bright
Source Sample (BSS; Massardi et al. 2008) covers the declination
region $\delta<-15^\circ$ and is complete in total intensity down to
$S_{20\rm GHz}=0.5\,$Jy (except for Fornax A) and has simultaneous
polarization measurements. Nine extended (i.e. with size larger than
the 2.4 arcmin resolution of ATCA measurements) objects have been
followed-up in polarization, mosaicking a region large enough to
evaluate reasonably well the integrated polarized flux density
\citep{bur09}. Despite all efforts, for some very extended objects
(like Cen A) the full extent of the low-frequency radio structure
could not be imaged and the total $P$ could not be measured. Angular
resolution plays a key role in the comparison of the measurements of
$P$ obtained with different instrument or configurations for extended
sources. The problems are, of course, amplified if observations at
different frequencies are compared: the AT20G observations showed that
the polarization degree may vary with frequency.  The BSS comprises
218 detections of polarized flux density, of which 19 are above 100
mJy at 20 GHz. 73 objects of our 95.00\% confidence level detections
are in common with the BSS, but only 28 of them have a polarization
detection in the AT20G BSS, with polarized flux density typically
lower than those estimated with the techniques applied here, as
detailed later.

Ricci et al. (2004) have carried out 18 GHz polarization observations
of the Southern portion of the \cite{kuh81} sample, comprising sources
with $S_{5\rm GHz}=1\,$Jy. Due to resolution effects, a source in the
Centaurus A region appears with flux density larger than 1 Jy in our
sample but with less than 100 mJy in the \cite{ric04} sample.
Variability may justify the disagreement for the blazar 3C\,279 for
which Ricci et al. measured $P_{18\rm GHz}=1.6\,$Jy, a factor of
$\simeq 2.5$ higher than estimated from the WMAP 23GHz map.

Both the BSS and the Ricci et al. samples cover the Southern
hemisphere. Among the Northern 78 VLA polarization calibrators listed
in the compilation updated by
S. Meyer\footnote{www.aoc.nrao.edu/~smyers/calibration/master.shtml}
31 are among our 95.00\% significance level detections. 13 of them
have polarized flux density (averaged over the epoch 2002-2006) above
100 mJy in the K band.

As mentioned above, the comparison with ground based measurements is
complicated by resolution effects for the extended (generally
steep-spectrum) sources and by the strong variability of the
flat-spectrum ones. Nevertheless, Fig.~\ref{fig:comparisons} shows a
reasonably good consistency for $P>400\,$mJy at 23 GHz.  Below 400
mJy, the values of $P$ are clearly dominated by the contribution of
positive polarization fluctuations at the source positions (Eddington
bias).

\subsection{Additional tests and simulations}\label{sec:sim}

Unfortunately the number of objects with polarized flux above 400 mJy
and ground based polarization measurement at frequencies close to WMAP
ones is small, so that our comparison has a poor statistics. To better
assess the reliability of our flux estimates it is thus necessary to
resort to simulations. We have selected a sample of 738 positions with
$|b|>5^\circ$ and far from each other more than $4R_s$ (see the first
paragraph of \S\,\ref{sec:pfe}) and more than $2R_s$ away from each
object in the NEWPS\_5yr\_3s catalogue. These `blank' positions (free
of sources brighter than $3\sigma$ in total intensity) constitute our
control fields. We have then chosen 10 values of $P$, ranging from 0.2
to 2 Jy, with a constant step in $\log P$, i.e. $P=[0.2, 0.26, 0.33,
0.43, 0.56, 0.719, 0.928, 1.20, 1.55, 2.0]$. For each value of $P$ we
have injected a source with that polarized flux density in the
projected $Q$ and $U$ patches centered on 100 randomly chosen control
field positions. This was done randomly selecting $Q$ between $-P$ and
$P$ and setting $U=\pm \sqrt{P^2-Q^2}$, the sign of $U$ being again
chosen at random. The inserted source was convolved with the WMAP
symmetrized beam at 23 GHz. We avoided using the same patch more than
once, except when all the control fields were already used. In any
case the same patch was never used twice for the same value of $P$.

\begin{figure}
\begin{center}
\includegraphics[width=0.45\textwidth]{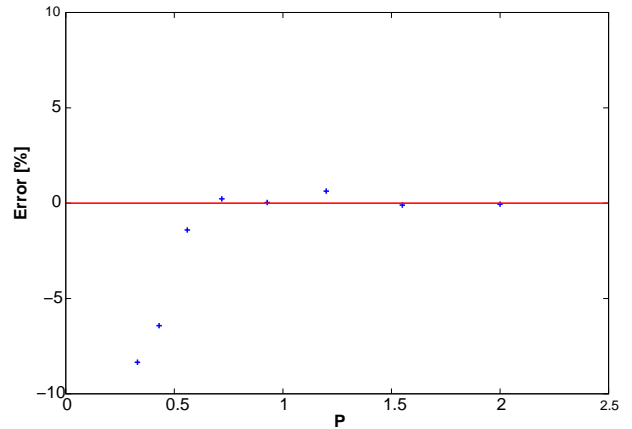}
\caption{Results of the simulations for logarithmically spaced values
  of $P$ in the range 0.2-2 Jy. In this panel we show the percentage
  error ($100(P_{\rm input}-P_{\rm recovered})/P_{\rm input}$) in the
  recovered value of $P$. \label{fig:P_Psim_2}}
\end{center}
\end{figure}

Figure~\ref{fig:P_Psim_2} shows the percentage error in the values of
$P$ of the simulated sources recovered with our filtering and flux
estimation process with respect to the input values.  This comparison
confirms the reliability of recovered fluxes for $P_{\rm input} \gsim
400$ mJy, while the Eddington bias becomes increasingly important
below this value. For $P_{\rm input} = 400$ mJy the recovered fluxes
are, on average, overestimated by less than 10\%.

\begin{figure}
\begin{center}
\includegraphics[width=0.5\textwidth]{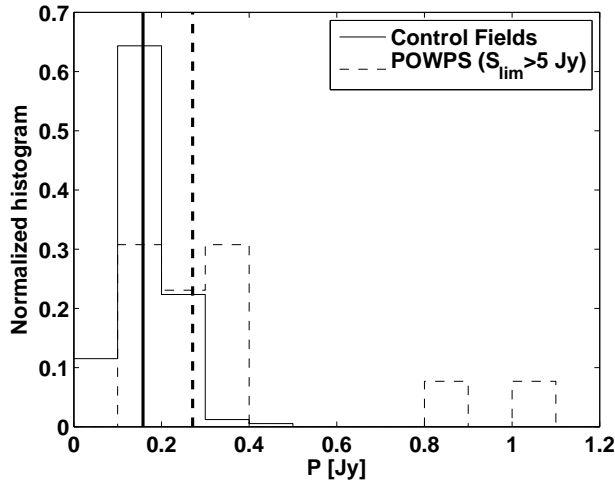}
\caption{Distributions, normalized to unity, of $P$ signals in the
  direction of the 23 GHz sources brighter than 5 Jy (13 objects;
  dashed histogram) compared with those for the control fields
  (solid). The median $P$ values are of 0.27 Jy for sources and of
  0.16 Jy for control fields. According to the Kolmogorov-Smirnov
  test, the probability that the two samples are drawn from the same
  parent distribution is of only $7.3\cdot 10^{-5}$ \label{fig:Phist}}
\end{center}
\end{figure}

\subsection{Median Polarization Degree}

The median polarization degree at 23 GHz of the 11 sources detected
with a confidence level $\ge 99.99\%$, with low radio frequency
counterparts, and with $P\ge 400\,$mJy is $\simeq 7.5\%$. This value,
however, is not representative of the mean polarization level of
sources in a complete sample for two reasons. First, the sample is
obviously biased towards sources with the highest polarization
degrees. Second, as mentioned above, our photometry is not optimal for
extended sources which make up a substantial fraction of the sample,
so that their estimated polarization degrees are highly uncertain.

An unbiased estimate of the mean polarization degree of sources at
WMAP frequencies can be obtained from a comparison of the distribution
of $P$ values for a suitably chosen complete subsample with that of
control fields. The completeness limit, in total flux, cannot be too
faint, otherwise the mean polarization level is too low to be
detectable. On the other hand, if the flux limit is too high, the
number of sources is too small for a meaningful statistical
inference. The optimal flux limits turn out to be of 5, 4, 3, and 4 Jy
at 23, 33, 41, and 61 GHz, respectively. We find a highly significant
detection of polarized flux density only at 23 GHz (see
Fig.~\ref{fig:Phist}): the probability that the distribution of $P$
values of sources is drawn from the same parent distribution as
control fields is $7.3\cdot 10^{-5}$. The median polarization degree,
$\Pi$, can be estimated as:
\begin{equation} \label{eq:PolDeg}
  \Pi = {P_{\rm med, sources}- P_{\rm med, control fields}\over S_{\rm med, sources}},
\end{equation}
where $S_{\rm med, sources}$ is the median flux of sources and we have
neglected the median flux of control fields, which is close to zero. The result
at 23 GHz is $\Pi_{23\rm GHz}= 1.7\pm 1.1$\%, consistent with the median
polarization degree for the AT20G BSS, $\Pi_{20\rm GHz}\simeq 2.5$\%
\citep{mas08}. Note that the lower resolution of WMAP, compared to AT20G,
observations make them more liable to beam depolarization (due to chaotic
components of the magnetic field within the unresolved region) in the case of
extended sources.

The probabilities that the distributions of $P$ values of sources are drawn
from the same parent distribution as control fields are of 0.02, 0.012, and
0.065  at 33, 41, and 61 GHz, respectively. The derived values of $\Pi$ are
$\Pi_{33\rm GHz}= 0.91\pm 0.83$\%, $\Pi_{41\rm GHz}= 0.68\pm 1.0$\%, and
$\Pi_{61\rm GHz}= 1.3\pm 1.8$\%.

\subsection{Predictions for Planck}

The Planck satellite is expected to measure the polarization of the
sources with greater sensitivity than WMAP, and therefore to detect
more sources, down to fainter polarized flux limits. Building on the
results of our analysis of WMAP maps, we have carried out simulations
to estimate the minimum polarized flux detectable and reliably
measurable by Planck. We have adopted the nominal mean instrumental
noise levels expected after two complete sky surveys (1
yr)\footnote{www.rssd.esa.int/index.php?project=Planck}.

Assuming that the $Q$ and $U$ images are dominated by instrumental
white noise, and adopting a pixel size of $6.87$ arcmin and idealized
matched filters for the nominal beam sizes, we have computed the
ratios between the $\sigma_P$ levels for the 30--100 GHz Planck
frequency channels and those for the closest WMAP frequency
channels. We find $\sigma_{P,\rm WMAP\,5yr}/\sigma_{P,\rm
  Planck\,1yr}= 2.2$, 1.6, 2.0, and 6.8 at about 30, 44, 70, and 100
GHz, respectively. Note that the higher Planck sensitivity is partly
compensated by the longer WMAP exposure time. An extension of the
Planck mission for one more year would decrease $\sigma_{P,\rm
  Planck}$ by a factor $\sqrt{2}$.

The above calculations take into account only instrumental noise. To
investigate the effect also of polarization fluctuations due to
diffuse foregrounds and to the CMB we have performed simulations
analogous to those we did for WMAP at 23 GHz for the Planck 30 GHz
channel, using the Planck Sky Model simulation software \citep{del09}.

In simulated $Q$ and $U$ maps containing polarized diffuse
foregrounds and the CMB we have injected sources with $P$ in the range
0.05--0.6 Jy. Our algorithm recovered sources with $P$ down to 300 mJy
with systematic offsets due to the Eddington bias of few percent or
less.

This result already highlights the difficulty of finding suitable
polarization calibrators for Planck. Not many point-like sources have
$P$ substantially larger than 300 mJy at 30 GHz. Combining the results
by Ricci et al. (2004), of the AT20G BSS (Massardi et al. 2008) and
the VLA calibrator observations, we have found 11 extragalactic
sources above this limit. The count may be incomplete, particularly in
the Northern hemisphere, but it is unlikely that many bright sources
have been missed. Our analysis of 33 GHz WMAP maps, including bright
Galactic sources, has detected 12 sources at a confidence level
$>99.99\%$; polarized flux densities for most of them, however, are
overestimated because of the Eddington bias. Also, some of the
brightest sources are extended (compared to the Planck beam) and
others (3C273 and 3C279) are highly variable.

\section{Conclusions} \label{sec:Conclusions}

We have applied to the WMAP 5-yr polarization maps a new source
detection technique, called ``filtered fusion'' (Arg\"ueso et
al. 2009), taking into account the real symmetrized beam profiles. The
technique was applied at the positions of WMAP sources detected at
$\ge 5\sigma$ by Massardi et al. (2009), plus 3 extended sources
(Cygnus A, Taurus A, Cas A) known to be bright and polarized. 22
sources were detected at a confidence level $\ge 99.99\%$ (that would
correspond to $\simeq 3.72\sigma$ for a one-tailed Gaussian
statistics) in at least one WMAP channel. Five of them, however, have
no plausible counterparts in low radio frequency catalogs and may
therefore be just high intensity peaks of the fluctuation
field. Nevertheless, this is a substantial improvement compared to the
5 source polarization measurements listed by \cite{wri09} and our
results for these 5 sources are in generally good agreement with
theirs.There are several reasons that could explain such an
improvement. First, WMAP used a mask to produce the point source
catalogs based on the KQ75 mask plus the Magellanic cloud regions that
exclude 11 objects that we have studied. Second, WMAP did not consider
objects that are not associated to known sources at low
frequencies. In our case, we do a non-blind search at the positions of
the NEWPS-5yr catalog that includes a handful of such objects, some of
which show a significant polarized flux (marked as NC in Table
\ref{tab:powps}), maybe due to Galactic emission in
polarization. Taking all these into account, and considering the
efficiency of our method as compared to WMAP, one could conclude that
we are detecting between $30-40\%$ more sources, which is a similar
improvement to the one obtained in our analysis in total intensity
\citep{mas08}.

A comparison of our polarized flux, $P$, estimates at 23 GHz
(Fig.~\ref{fig:comparisons}, where sources detected at $\ge 99.99\%$
confidence level are represented by filled symbols) with high
signal-to-noise ground based measurements at nearby frequencies,
highlights the complications due to different angular resolutions in
the case of extended sources (Cen A and Vir A) and of strong
variability (3C279). For sources not affected by these problems, the
agreement is quite good. The latter sources, however, are too few to
allow a firm conclusion on the reliability of our flux estimate. We
have therefore resorted to simulations, injecting fake sources of
known polarized flux density in the WMAP 23 GHz $Q$ and $U$ maps. The
simulations showed that our approach is reliable for $P_{23\rm GHz}
\ge 400\,$mJy, while the Eddington bias becomes increasingly large at
fainter fluxes (it is less than $10\%$ at
400\,mJy). Figure~\ref{fig:comparisons} also shows that estimates of
$P$ for sources detected at a confidence level $< 99.99\%$ are badly
affected by the Eddington bias, and therefore unreliable. This was
expected since, for example, a 99.90\% confidence level corresponds to
$3.1\sigma$ in the case of a Gaussian distribution of
fluctuations. And, as shown by Hogg \& Turner (1998), flux estimates
at this confidence level are practically useless.

Analogous simulations using the mean noise levels expected after 1
year of Planck observations have shown that a few percent Eddington
bias is reached for $P_{30\rm GHz} \simeq 300\,$mJy. The modest
difference compared to WMAP 23 GHz is due to the fact that the longer
WMAP exposures (5 yrs vs. 1 yr) compensate for the lower
sensitivity. In the case of an extension of the Planck mission for a
second year, the gain over WMAP will increase by a factor of
$\sqrt{2}$. Based on the predictions by Taylor et al. (2007; their
Fig.~7), this would imply an increase of detections in polarization by
a factor $\gsim 3$ compared to WMAP.

Sources detected on WMAP polarization maps have, on average,
exceptionally high polarization degrees because only a very small
fraction of sources have been detected and they likely populate the
tail of the distribution of P values. Estimates of, or upper limits
to, the mean polarization degrees, $\Pi$, of bright sources at 23, 33,
41, and 61 GHz have been obtained comparing the distributions of
polarized flux densities in the directions of complete source samples,
limited in total flux, with those in the directions of control fields,
devoid of bright sources. The derived values, $\Pi_{23\rm GHz}= 1.7\pm
1.1$\%, $\Pi_{33\rm GHz}= 0.91\pm 0.83$\%, $\Pi_{41\rm GHz}= 0.68\pm
1.0$\% and $\Pi_{61\rm GHz}= 1.3\pm 1.8$\%, are consistent with, but
somewhat on the low-side of, the median polarization degree for the
AT20G BSS, $\Pi_{20\rm GHz}\simeq 2.5$\% \citep{mas08}. Somewhat lower
values of $\Pi$ are expected at the WMAP resolution for extended
sources, due to beam depolarization.

Finally, the detected sources may be candidate calibrators for other
high sensitivity CMB experiments, like the ground-based QUIJOTE
experiment (Rubi\~no-Martin et al. 2008) or the Planck satellite
mission.

\section*{Acknowledgements}
MLC acknowledges a postdoctoral fellowship from the Spanish MEC in
Cambridge (UK) and an EGEE-III postdoctoral contract at IFCA. JGN
acknowledges a researcher position grant at the SISSA-ISAS
(Trieste). Partial financial support for this research has been
provided to MM, JGN and GDZ by the Italian ASI (contracts Planck LFI
Activity of Phase E2 and I/016/07/0 'COFIS'), and to JLS by the
Spanish MEC. LL acknowledges a JAE-predoc fellowship from the Spanish
CSIC. The authors acknowledge the use of the Planck Sky Model,
developed by the Component Separation Working Group (WG2) of the
Planck Collaboration. The authors wish to thank Bruce Partridge for
useful discussions during the preparation of this work.

\newpage


\begin{thebibliography}{}

\bibitem[\protect\citeauthoryear{Arg{\"u}eso et al.}{2009}]{arg09}
  Arg{\"u}eso F., Sanz J.L., Herranz D., L{\'o}pez-Caniego M.,
  Gonz{\'a}lez-Nuevo, J., 2009, MNRAS, 395, 649

\bibitem[\protect\citeauthoryear{Barnard et
al.}{2004}]{bar04} Barnard V.~E., Vielva P., Pierce-Price
D.~P.~I., Blain A.~W., Barreiro R.~B., Richer J.~S., Qualtrough C., 2004,
MNRAS, 352, 961

\bibitem[\protect\citeauthoryear{Bock et al.}{2006}]{boc06}
Bock J. et al., 2006, arXiv:astro-ph/0604101

\bibitem[\protect\citeauthoryear{Burke-Spolaor et al.}{2009}]{bur09} Burke-Spolaor S., Ekers R.~D., Massardi M.,
Murphy T., Partridge B., Ricci R., Sadler E.~M., 2009, MNRAS, 395, 504

\bibitem[\protect\citeauthoryear{Carvalho, Rocha,
\& Hobson}{2009}]{car09} Carvalho P., Rocha G., Hobson M.~P., 2009, MNRAS, 393, 681

\bibitem[\protect\citeauthoryear{Delabrouille et al.}{2009}]{del09} Delabrouille J. et al., 2009, in preparation.

\bibitem[\protect\citeauthoryear{Eddington}{1913}]{1913MNRAS..73..359E}
Eddington A.~S., 1913, MNRAS, 73, 359

\bibitem[\protect\citeauthoryear{Feroz \& Hobson}{2008}]{fer08} Feroz F., Hobson M.~P., 2008, MNRAS, 384, 449

\bibitem[\protect\citeauthoryear{Gonz{\'a}lez-Nuevo et
    al.}{2006}]{gon06} Gonz{\'a}lez-Nuevo J., Arg{\"u}eso F.,
  L{\'o}pez-Caniego M., Toffolatti L., Sanz J.~L., Vielva P., Herranz
  D., 2006, MNRAS, 369, 1603

\bibitem[\protect\citeauthoryear{Gonz{\'a}lez-Nuevo et
al.}{2008}]{2008MNRAS.384..711G} Gonz{\'a}lez-Nuevo J., Massardi M., Arg{\"u}eso F., Herranz D., Toffolatti L., Sanz
J.~L., L{\'o}pez-Caniego M., de Zotti G., 2008, MNRAS, 384, 711

\bibitem[\protect\citeauthoryear{G{\'o}rski et al.}{2005}]{gor05}
  G{\'o}rski K.~M., Hivon E., Banday A.~J., Wandelt B.~D., Hansen
  F.~K., Reinecke M., Bartelmann M., 2005, ApJ, 622, 759

\bibitem[\protect\citeauthoryear{Herranz et al.}{2002}]{her02}
  Herranz D., Sanz J.~L., Barreiro R.~B.,
  Mart{\'{\i}}nez-Gonz{\'a}lez E., 2002, ApJ, 580, 610

\bibitem[\protect\citeauthoryear{Herranz et
al.}{2009}]{her08a} Herranz D., L{\'o}pez-Caniego M., Sanz
J.~L., Gonz{\'a}lez-Nuevo J., 2009, MNRAS, 394, 510

\bibitem[\protect\citeauthoryear{Herranz
\& Sanz}{2008}]{her08b} Herranz D., Sanz J.~L., 2008, JSTSP, 5, 727

\bibitem[\protect\citeauthoryear{Hobson
\& McLachlan}{2003}]{hob03} Hobson M.~P., McLachlan C., 2003, MNRAS, 338, 765

\bibitem[\protect\citeauthoryear{Hogg
\& Turner}{1998}]{1998PASP..110..727H} Hogg D.~W., Turner E.~L., 1998, PASP, 110, 727

\bibitem[\protect\citeauthoryear{Homan, Attridge \&
    Wardle}{2001}]{hom01} Homan D.~C., Attridge J.~M., Wardle J.~F.~C.,
  2001, ApJ, 556, 113

\bibitem[\protect\citeauthoryear{Kamionkowski, Kosowsky
\& Stebbins}{1997}]{kam97} Kamionkowski M., Kosowsky A., Stebbins A., 1997, PhRvD, 55, 7368

\bibitem[\protect\citeauthoryear{Kuehr et al.}{1981}]{kuh81} Kuehr H., Witzel A., Pauliny-Toth I.~I.~K., Nauber U., 1981, A\&AS, 45, 367 

\bibitem[\protect\citeauthoryear{Leach et
al.}{2008}]{lea08} Leach S.~M., et al., 2008, A\&A, 491, 597

\bibitem[\protect\citeauthoryear{L{\'o}pez-Caniego et
  al.}{2005}]{lop05a} L{\'o}pez-Caniego M., Herranz D., Barreiro R.B.,
  Sanz J.L., 2005a, MNRAS, 359, 993

\bibitem[\protect\citeauthoryear{L{\'o}pez-Caniego et
al.}{2005}]{lop05b} L{\'o}pez-Caniego M., Herranz D., Sanz J.~L., Barreiro R.~B., 2005b, EJASP, 2005, 2426

\bibitem[\protect\citeauthoryear{L{\'o}pez-Caniego et
  al.}{2006}]{lop06} L{\'o}pez-Caniego M., Herranz D.,
  Gonz{\'a}lez-Nuevo J., Sanz J.~L., Barreiro R.~B., Vielva P.,
  Arg{\"u}eso F., Toffolatti L., 2006, MNRAS, 370, 2047

\bibitem[\protect\citeauthoryear{L{\'o}pez-Caniego et
al.}{2007}]{lop07} L{\'o}pez-Caniego M., Gonz{\'a}lez-Nuevo
J., Herranz D., Massardi M., Sanz J.~L., De Zotti G., Toffolatti L.,
Arg{\"u}eso F., 2007, ApJS, 170, 108

\bibitem[\protect\citeauthoryear{Malik
\& Subramanian}{1997}]{mal97} Malik R.~K., Subramanian K., 1997, A\&A, 317, 318


\bibitem[\protect\citeauthoryear{Massardi et
al.}{2008}]{mas08} Massardi M., et al., 2008, MNRAS, 384, 775

\bibitem[\protect\citeauthoryear{Massardi et
al.}{2009}]{mas09} Massardi M., L{\'o}pez-Caniego M.,
Gonz{\'a}lez-Nuevo J., Herranz D., de Zotti G., Sanz J.~L., 2009, MNRAS,
392, 733

\bibitem[\protect\citeauthoryear{Nailong}{1992}]{nai92}
Nailong W., 1992, ASPC, 25, 291

\bibitem[\protect\citeauthoryear{Ricci et al.}{2004}]{ric04}
  Ricci R., et al., 2004, MNRAS, 354, 305

\bibitem[\protect\citeauthoryear{Rubino-Martin et
al.}{2008}]{2008arXiv0810.3141R} Rubi\~no-Martin J.~A., et al., 2008, arXiv:0810.3141

\bibitem[\protect\citeauthoryear{Sadler et al.}{2008}]{sad08}
Sadler E.~M., Ricci R., Ekers R.~D., Sault R.~J., Jackson C.~A., de Zotti
G., 2008, MNRAS, 385, 1656

\bibitem[\protect\citeauthoryear{Sanz, Herranz \&
  Mart{\'{\i}}nez-G{\'o}nzalez}{2001}]{san01} Sanz J.~L.,
  Herranz D., Mart{\'{\i}}nez-G{\'o}nzalez E., 2001, ApJ, 552, 484
%

\bibitem[\protect\citeauthoryear{Sanz et al.}{2006}]{san06}
Sanz J.~L., Herranz D., L{\'o}pez-Caniego M., Argueso F., 2006, arXiv:astro-ph/0609351

\bibitem[\protect\citeauthoryear{Stewart}{2006}]{ste06} Stewart I.~M., 2006, A\&A, 454, 997
%

\bibitem[\protect\citeauthoryear{Taylor et al.}{2007}]{2007ApJ...666..201T}
Taylor A.~R., et al., 2007, ApJ, 666, 201

\bibitem[\protect\citeauthoryear{Tegmark
\& de Oliveira-Costa}{1998}]{teg98} Tegmark M., de Oliveira-Costa A., 1998, ApJ, 500, L83

\bibitem[\protect\citeauthoryear{Tucci et al.}{2004}]{tuc04}
Tucci M., Mart{\'{\i}}nez-Gonz{\'a}lez E., Toffolatti L.,
Gonz{\'a}lez-Nuevo J., De Zotti G., 2004, MNRAS, 349, 1267

\bibitem[\protect\citeauthoryear{Tucci et al.}{2005}]{tuc05}
Tucci M., Mart{\'{\i}}nez-Gonz{\'a}lez E., Vielva P., Delabrouille J.,
2005, MNRAS, 360, 935

\bibitem[\protect\citeauthoryear{Vielva et al.}{2001}]{vie01}
  Vielva P., Mart{\'{\i}}nez-Gonz{\'a}lez E., Cay{\'o}n L., Diego J.~M., Sanz
  J.~L., Toffolatti L., 2001, MNRAS, 326, 181
%
\bibitem[\protect\citeauthoryear{Vielva et al.}{2003}]{vie03}
  Vielva P., Mart{\'{\i}}nez-Gonz{\'a}lez E., Gallegos J.~E., Toffolatti L.,
  Sanz J.~L., 2003, MNRAS, 344, 89

\bibitem[\protect\citeauthoryear{Vikhlinin et
al.}{1995}]{vik95} Vikhlinin A., Forman W., Jones C., Murray
S., 1995, ApJ, 451, 542

\bibitem[\protect\citeauthoryear{Wright et al.}{2009}]{wri09}
Wright E.~L., et al., 2009, ApJS, 180, 283

\end{thebibliography}
\end{document}